\documentclass[sigconf]{acmart}
\AtBeginDocument{%
  }

\acmConference[EASE 2025]{The 29th International Conference on Evaluation and Assessment in Software Engineering}
\usepackage{multirow}
\begin{document}

%%
%% The "title" command has an optional parameter,
%% allowing the author to define a "short title" to be used in page headers.
\title{Exploring Zero-Shot App Review Classification with ChatGPT: Challenges and Potential}

%%
%% The "author" command and its associated commands are used to define
%% the authors and their affiliations.
%% Of note is the shared affiliation of the first two authors, and the
%% "authornote" and "authornotemark" commands
%% used to denote shared contribution to the research.

\author{Mohit Chaudhary}
\email{mohit.chaudhary3@tcs.com}
\affiliation{%
  \institution{TCS Research}
  \city{Pune}
  \country{India}}

\author{Chirag Jain}
\email{chirag.rjain3@tcs.com}
\affiliation{%
  \institution{TCS Research}
  \city{Pune}
  \country{India}}

\author{Preethu Rose Anish}
\email{preethu.rose@tcs.com}
\affiliation{%
  \institution{TCS Research}
  \city{Pune}
  \country{India}}

%%
%% By default, the full list of authors will be used in the page
%% headers. Often, this list is too long, and will overlap
%% other information printed in the page headers. This command allows
%% the author to define a more concise list
%% of authors' names for this purpose.
\renewcommand{\shortauthors}{Mohit et al.}

%%
%% The abstract is a short summary of the work to be presented in the
%% article.
\begin{abstract}
  App reviews are a critical source of user feedback, offering valuable insights into an app's performance, features, usability, and overall user experience. Effectively analyzing these reviews is essential for guiding app development, prioritizing feature updates, and enhancing user satisfaction. Classifying reviews into functional and non-functional requirements play a pivotal role in distinguishing feedback related to specific app features (functional requirements) from feedback concerning broader quality attributes, such as performance, usability, and reliability (non-functional requirements). Both categories are integral to informed development decisions. Traditional approaches to classifying app reviews are hindered by the need for large, domain-specific datasets, which are often costly and time-consuming to curate. This study explores the potential of zero-shot learning with ChatGPT for classifying app reviews into four categories: functional requirement, non-functional requirement, both, or neither. We evaluate ChatGPT's performance on a benchmark dataset of 1,880 manually annotated reviews from ten diverse apps spanning multiple domains. Our findings demonstrate that ChatGPT achieves a robust F1 score of 0.842 in review classification, despite certain challenges and limitations. Additionally, we examine how factors such as review readability and length impact classification accuracy and conduct a manual analysis to identify review categories more prone to misclassification.
\end{abstract}

%%
%% The code below is generated by the tool at http://dl.acm.org/ccs.cfm.
%% Please copy and paste the code instead of the example below.
%%
% \begin{CCSXML}
% <ccs2012>
%  <concept>
%   <concept_id>00000000.0000000.0000000</concept_id>
%   <concept_desc>Do Not Use This Code, Generate the Correct Terms for Your Paper</concept_desc>
%   <concept_significance>500</concept_significance>
%  </concept>
%  <concept>
%   <concept_id>00000000.00000000.00000000</concept_id>
%   <concept_desc>Do Not Use This Code, Generate the Correct Terms for Your Paper</concept_desc>
%   <concept_significance>300</concept_significance>
%  </concept>
%  <concept>
%   <concept_id>00000000.00000000.00000000</concept_id>
%   <concept_desc>Do Not Use This Code, Generate the Correct Terms for Your Paper</concept_desc>
%   <concept_significance>100</concept_significance>
%  </concept>
%  <concept>
%   <concept_id>00000000.00000000.00000000</concept_id>
%   <concept_desc>Do Not Use This Code, Generate the Correct Terms for Your Paper</concept_desc>
%   <concept_significance>100</concept_significance>
%  </concept>
% </ccs2012>
% \end{CCSXML}

% \ccsdesc[500]{Do Not Use This Code~Generate the Correct Terms for Your Paper}
% \ccsdesc[300]{Do Not Use This Code~Generate the Correct Terms for Your Paper}
% \ccsdesc{Do Not Use This Code~Generate the Correct Terms for Your Paper}
% \ccsdesc[100]{Do Not Use This Code~Generate the Correct Terms for Your Paper}

%%
%% Keywords. The author(s) should pick words that accurately describe
%% the work being presented. Separate the keywords with commas.
\keywords{App Review Classification, Zero-Shot Learning,
ChatGPT, Prompt Engineering, Requirement Engineering}
%% A "teaser" image appears between the author and affiliation
%% information and the body of the document, and typically spans the
%% page.
% \begin{teaserfigure}
%   \includegraphics[width=\textwidth]{sampleteaser}
%   \caption{Seattle Mariners at Spring Training, 2010.}
%   \Description{Enjoying the baseball game from the third-base
%   seats. Ichiro Suzuki preparing to bat.}
%   \label{fig:teaser}
% \end{teaserfigure}

% \received{20 February 2007}
% \received[revised]{12 March 2009}
% \received[accepted]{5 June 2009}

%%
%% This command processes the author and affiliation and title
%% information and builds the first part of the formatted document.
\maketitle

\section{Introduction}
In the digital age, mobile applications (apps) have become integral to everyday life, serving a wide range of functions across various industries. As the demand for high-quality apps continues to grow, understanding and addressing user feedback is crucial for developers to improve app quality and enhance user satisfaction. App reviews offer valuable insights into user experiences, preferences, and unmet needs, providing critical feedback that identifies improvement opportunities and informs the development process through aggregated user knowledge \cite{9} \cite{18}.

Requirement gathering is a critical phase in the software development lifecycle, during which  the needs and expectations of users are identified \cite{rw4} . The success of a software system is largely determined by how well it meets user needs and adapts to its operating environment \cite{laghari}. Consequently, accurate and thorough requirements gathering is essential for delivering a high-quality software product. However, identifying and classifying user reviews that contain actionable requirements can be particularly challenging due to the unstructured nature of user feedback and the diverse ways in which users express their opinions \cite{1}.

User reviews generally address two main types of requirements: functional requirements (FRs) and non-functional requirements (NFRs). FRs typically focus on core application functionality, such as bugs in critical features like payment processing, which can shape the development roadmap. NFRs are typically associated with the system performance, including crashes, lag, or performance degradation, which affect reliability, user satisfaction, and experience. Both are crucial for user retention and overall app quality.

Effectively identifying and classifying user reviews into FRs and NFRs can streamline the development process. Although bug reports and feature requests are invaluable, they often fail to encompass the full spectrum of user experiences with the app. Distinguishing between FRs and NFRs allows developers to better understand whether the feedback pertains to how the app functions (FRs) or how it performs (NFRs). This distinction is critical, as both categories affect development strategy and resource allocation in different ways.

FRs typically necessitate significant development efforts, such as building new features or correcting functional logic. In contrast, NFRs focus on improving system quality attributes, such as performance (speed, latency), usability, and security. By categorizing user feedback into these two types of requirements, the feedback can be directed to the appropriate teams, ensuring clearer communication among stakeholders and more effective allocation of resources. For instance, developers may focus on functional features, while designers and other specialists can address NFRs, such as optimizing performance or enhancing security. This clear delineation of responsibilities promotes more focused and efficient development processes, allowing each team to leverage their domain-specific expertise to enhance various aspects of the application.

Traditional machine learning models like Support Vector Machines (SVM), Decision Trees, and Random Forests have long been standard approaches for text classification. However, they rely heavily on manually labeled, domain-specific datasets, a process that is time-consuming, costly, and requires domain expertise. Additionally, these models struggle to capture context and nuances in language, limiting their ability to understand sentiment and intent \cite{26}. In contrast, Large Language Models (LLMs), such as ChatGPT, have emerged as powerful deep learning tools, excelling in tasks like text classification. ChatGPT’s “zero-shot” learning capability enables it to perform classifications without extensive fine-tuning on domain-specific datasets, making it particularly useful for tasks like app review classification \cite{hamid} \cite{israt}. This approach enables models to leverage general language understanding to classify reviews into distinct requirement types, eliminating the need for pre-labeled training sets and reduces the need for resource-intensive training sets, offering a cost-effective, time-efficient solution in dynamic environments like the app market, where review data is constantly evolving.

In this study, we explore various prompting techniques to classify app reviews into four categories: FRs, NFRs, both, or neither. We also assess the performance of the zero-shot GPT 4o mini model in this classification task, examining both its benefits and potential challenges. To evaluate the model’s performance, we constructed a benchmark dataset consisting of 1,880 manually annotated app reviews from ten diverse applications.

Our investigation is guided by the following research questions:

\begin{itemize}
    \item \textbf{RQ1:} How accurately does ChatGPT classify app review requirements into FRs, NFRs, both, or neither, and how does its performance compare to traditional ML models?
    \item \textbf{RQ2:} How do the length and level of detail in app reviews impact ChatGPT’s classification performance?
    \item \textbf{RQ3:} Are certain types of FRs or NFRs more prone to misclassification?
\end{itemize}

Preliminary results indicate that ChatGPT performs reasonably well in categorizing app reviews, achieving an F1 score of 0.842. However, our analysis also reveals several challenges and limitations when using ChatGPT for this task. We explore the effects of factors such as review readability and length on classification accuracy. Additionally, a manual analysis identifies specific types of reviews that are more susceptible to misclassification.

The remainder of the paper is structured as follows: Section 2 details the methodology and dataset used in the study. Section 3 provides an analysis of the model’s performance, and the challenges faced. Section 4 addresses threats to validity. Section 5 reviews related work, and Section 6 concludes the paper and highlighting potential areas for future research.

\section{Methodology}

In this section, we detail the dataset and the curation process, outline the method for selecting prompts through various prompt designs and techniques, and provide information on the zero-shot model used, including its parameters.

\subsection{Dataset}

We compiled a dataset of 2,000 app reviews from both the Google Play Store and Apple app store, representing ten diverse application domains: Communication (WhatsApp), Travel (Uber), Music \& Audio (Spotify), Social Media (Twitter), Video Player \& Editor (YouTube), Entertainment (Netflix), Games (Candy Crush Saga), Shopping (Amazon), Education (Duolingo), and Health (Google Fit). The reviews were selected based on the following criteria: (1) a minimum length of 10 words, (2) written in English, and (3) drawn from the most recent user feedback. 
To categorize the reviews, a team of five software engineers, each with over five years of professional experience in app development, manually annotated the reviews into one of the four classes: FRs, NFRs, both, or neither. To assess inter-annotator consistency, we computed Fleiss’s kappa score, which yielded a substantial level of agreement of 0.76. In cases where full agreement was not reached, a reconciliation process was initiated. Reviews with a majority consensus (i.e. at least three annotators in agreement) were retained in the final dataset. For reviews where significant disagreement remained, further discussions were held to resolve differences. As a result of this process, 120 reviews that could not reach consensus were excluded, leading to a final curated dataset of 1,880 app reviews with four classes FR (712), NFR (654), Both (298) and Neither (216). This dataset served as a basis for evaluating the performance of the classification model.

% \vspace{-0.5\baselineskip} 
\subsection{Prompt Design}

We conducted a series of experiments to evaluate the efficacy of various prompting techniques for classifying app reviews into FRs, NFRs, both or neither. The prompting techniques tested include Emotion Prompting \cite{emotion}, Role Prompting \cite{role} and  Chain of Thought (CoT) \cite{cot}. The objective was to identify the most effective strategy for accurately classifying app reviews into specific requirements.

In our initial experiment, we applied eight distinct prompts to a sample of 100 reviews drawn from the curated dataset of 1,880 reviews. Each prompt incorporated unique characteristics and instructions with different prompting techniques. These prompts were evaluated based on their ability to accurately classify reviews into the specified categories. After this preliminary evaluation, we selected the top three performing prompts for further analysis on the complete dataset. Prompt which combined Emotion Prompting, Role Prompting, and Chain of Thought (CoT), yielded the most promising results for the classification task. A summary of the three prompts, along with details of their respective techniques and characteristics, is provided in Table \ref{table1}.

All three selected prompts were framed as human messages, utilizing a consistent system prompt: “Imagine you are an expert requirements analyst specializing in the complex classification of app reviews into one of four categories: FRs, NFRs, both, or neither. Provide only the category—no explanations.” This prompt was specifically designed to implement role prompting technique. By combining role assignments with reasoning tasks within a single message, this approach facilitates streamlined, single-turn interaction when querying ChatGPT.

\begin{table}[htbp]
\caption{Prompts Description}
\begin{center}
% \scriptsize
\footnotesize
\begin{tabular}{|p{1.2cm}|p{3.2cm}|p{3.2cm}|}
\hline \textbf{Prompt} &  \textbf{Prompt characteristics} & \textbf{Prompt content} 
\\ \hline

Prompt 1 & A detailed prompt that provides comprehensive definitions of each type of requirement, resulting in a longer format with utilization of role prompting.  & Analyze the following app review and categorize it as relating to functional requirements, non-functional requirements, both, or neither, based on the definitions below:

Functional requirements: They specify what a system should do. They describe the specific behaviors, functions, and features of the application or system. Essentially, these are the \textquotedblleft what" the system should accomplish.

Non-functional requirements: They describe how a system should perform a function rather than what it should do. They focus on the quality attributes, such as performance, usability, reliability, and security.

Both: If review contains both functional and non-functional requirements.
None: If review does not contain any type of requirement.
     \\ \hline
Prompt 2 & A clear, concise prompt in statement form that avoids definitions, while implementing a zero-shot CoT and role prompting approach. & Read the app review below and decide whether it discusses functional requirements, non-functional requirements, both, or none. Let’s think this out in step by step way to be sure we have the right answer.      
 \\ \hline

Prompt 3 & A prompt that includes brief definitions of the requirement types with utilization of ensemble of emotion prompting, CoT and role prompting approach. & Classify the following app review into one of the following categories: functional requirements (features or functionalities needed), non-functional requirements (performance, usability, etc.), both (if it mentions both types of requirements), or none (if it doesn't fit into either category). Let’s think this out in step by step way to be sure we have the right answer. This is important to my career.       
 \\ \hline

\end{tabular}
\label{table1}
\end{center}
\end{table}

\subsection{Querying ChatGPT}
We employed Azure OpenAI’s GPT-4o mini to classify app reviews into specific requirement categories. Each review in the dataset was processed through a single API call, where the input consisted of a task prompt and the corresponding app review. Given that LLM outputs are inherently non-deterministic \cite{11}, we investigated the impact of varying temperature settings on the classification process. The three temperature settings (0.2, 0.7, and 1.3) were chosen to explore different levels of output randomness and creativity in the model's responses. These values represent a low, medium, and high level of randomness, respectively. Lower temperature values, such as 0.2, yield more deterministic and focused results, while higher temperature values (1.0 and above) introduce increased variability and flexibility which might be helpful in classifying more nuanced or ambiguous reviews. The mid-range value of 0.7 was included to observe the balance between creativity and reliability.

\section{Performance and challenges} 
In this section, we evaluate the performance of ChatGPT using precision, recall, and the micro F1 score across various prompting strategies and temperature settings. These metrics offer a comprehensive assessment of the model’s performance in multi-class classification tasks. Precision and recall measure the model’s ability to minimize false positives and accurately identify relevant instances for each class. The micro F1 score aggregates true positives, false positives, and false negatives across all classes, providing a balanced evaluation of performance, particularly in cases of imbalanced class distributions. Together, these metrics allow us to assess both individual class performance and overall classification effectiveness. We also discuss challenges encountered, with a particular focus on how the length and complexity of app reviews impact classification accuracy. Furthermore, we compare ChatGPT’s performance with that of traditional ML models to establish a benchmark and better understand its relative effectiveness.

\noindent \textbf{\textit{RQ1: How accurately does ChatGPT classify app review requirements into FRs, NFRs, both, or neither, and how does its performance compare to traditional ML models?}}

This RQ examines ChatGPT’s ability to perform multi-class classification on app reviews, categorizing them into four classes: FRs, NFRs, both, or neither. The results of the three selected prompting strategies, evaluated across different temperature settings (0.2, 0.7, 1.3), are presented in Table \ref{table2}. These results highlight the influence of both the prompt type and the temperature setting on classification performance.

Our analysis reveals that temperature settings significantly affect performance, with all three prompts yielding better results at a lower temperature of 0.2 compared to higher values. This finding emphasizes the importance of controlling randomness in the model’s responses to optimize classification accuracy.

Among the three prompts, Prompt 3 achieved the highest F1 score of 0.842 at a temperature of 0.2, outperforming all other prompts and temperature configurations. This suggests that Prompt 3 is particularly effective, likely due to its combination of role prompting, emotion prompting, and CoT prompting. Role prompting directs the model to adopt a specific perspective, enhancing contextual understanding, while CoT prompting encourages step-by-step reasoning, improving classification precision. Emotion prompting introduces emotional context, resulting in more thoughtful and nuanced responses. The synergy of these techniques allows the model to classify app reviews more effectively.

\begin{table}[htbp]
\caption{Performance on different prompts and temperatures}
\centering
\begin{tabular}{|c|c|c|c|c|}
\hline
\textbf{Temperature} & \textbf{Prompt} & \textbf{Recall} & \textbf{Precision} & \textbf{F1 score} \\ \hline
\multirow{3}{*}{0.2} & P1 & 0.796 & 0.803 & 0.784 \\ \cline{2-5}
                     & P2 & 0.786 & 0.791 & 0.777 \\ \cline{2-5}
                     & P3 & 0.848 & 0.858 & \textbf{0.842} \\ \hline
\multirow{3}{*}{0.7} & P1 & 0.732 & 0.739 & 0.725 \\ \cline{2-5}
                     & P2 & 0.781 & 0.792 & 0.773 \\ \cline{2-5}
                     & P3 & 0.743 & 0.750 & 0.735 \\ \hline
\multirow{3}{*}{1.3} & P1 & 0.670 & 0.669 & 0.664 \\ \cline{2-5}
                     & P2 & 0.658 & 0.656 & 0.650 \\ \cline{2-5}
                     & P3 & 0.668 & 0.666 & 0.660 \\ \hline
\end{tabular}
\label{table2}
\end{table}

Additionally, we conducted a comparative analysis of the model’s performance using its optimal configuration (Prompt 3 with a temperature of 0.2) against traditional ML models. We evaluated several ML classification models on the curated dataset, which was split into 1410 reviews for training and 470 reviews for testing. For embedding generation, we used the pretrained sentence transformer model ‘all-MiniLM-L6-v2' to convert each review into 384-dimensional embeddings. Sentence transformers offer notable advantages over traditional word embedding techniques, such as Word2Vec, for text classification, as they capture the overall meaning of a sentence, rather than focusing solely on individual word meanings \cite{elif}.

On the test set, the zero-shot ChatGPT model (with the optimal configurations) was compared to five classical ML models, as shown in Table \ref{3}. The results demonstrated that the zero-shot ChatGPT model outperformed all traditional classification models by a significant margin.

\begin{table}[htbp]
\centering
\caption{Comparison with traditional ML models}
\begin{tabular}{|c|c|c|c|} % 4 columns with vertical lines
\hline
\textbf{Model} & \textbf{Precision} & \textbf{Recall} & \textbf{F1 Score} \\ \hline
Random Forest & 0.6 & 0.52 & 0.45 \\ \hline
Decision Tree & 0.35 & 0.35 & 0.35 \\ \hline
Support Vector Classifier & 0.46 & 0.54 & 0.49 \\ \hline
XgBoost & 0.46 & 0.51 & 0.46 \\ \hline
Logistic Regression & 0.47 & 0.53 & 0.48 \\ \hline
GPT 4o mini & 0.85 & 0.83 & \textbf{0.82}\\ \hline
\end{tabular}
\label{3}
\end{table}

\noindent \textbf{\textit{RQ2: How do the length and level of detail in app reviews affect ChatGPT’s classification performance?}}

To assess the challenges and limitations of ChatGPT with an optimized configuration, we investigated how factors such as review length, and complexity influenced the model's classification performance. Our analysis revealed that review length has a minimal effect on the classification accuracy of app requirements. Specifically, the average character count for misclassified reviews was 402, compared to 400 for correctly classified reviews, indicating that review length does not significantly impact performance.

In contrast, review complexity, as measured by the Flesch Kincaid Grade Level (FKGL) metric \cite{12}, was found to substantially affect classification accuracy. The FKGL metric evaluates text readability based on sentence length and syllable count, with lower scores indicating simpler text. Reviews with lower FKGL scores were more likely to be correctly classified, suggesting that simpler, more straightforward expressions of need are easier for the model to interpret accurately. 

\begin{equation}
\text{FKGL} = 0.39 \frac{N_{\text{words}}}{N_{\text{sentences}}} + 11.8 \frac{N_{\text{syllables}}}{N_{\text{words}}} - 15.59
\end{equation}

Specifically, reviews that were correctly classified had an average FKGL score of 6.34, while misclassified reviews had a higher average FKGL score of 9.24, as shown in Table \ref{table44}. This finding indicates that reviews with more complex structures, including ambiguous phrasing, convoluted sentence structures, or dense vocabulary, present greater challenges for the model. Such complexity can obscure the intent of the review, leading to a decrease in classification accuracy. Our results underscore the importance of simplifying review language and structure to improve the model’s ability to accurately identify core requirements.
% \vspace{-0.5\baselineskip}
\begin{table}[htbp]
\centering
\begin{center}
\caption{Impact of review length and complexity on model performance}\label{table44}

\centering
\begin{tabular}{|p{2cm}|p{2.5cm}|p{2.4cm}|} % 2 columns with vertical lines
\hline
    &\textbf{Average FKGL score} & \textbf{Average Character Length} \\ \hline
Correctly classified reviews & 6.34 & 400 \\ \hline
Misclassified reviews & 9.24 & 402 \\ \hline
\end{tabular}
\end{center}
%\label{table4}
\end{table}

\noindent \textbf{\textit{RQ3: Are certain types of FRs or NFRs more prone to misclassification?}}

This RQ investigates whether certain categories of requirements are more prone to misclassification by the model. Using the model’s optimized configuration, we observed an F1 score of 0.91 for FRs and 0.87 for NFRs. The model demonstrated particularly strong performance in classifying reviews that do not contain explicit requirements but instead express user praise, achieving an impressive F1 score of 0.94. These reviews typically feature straightforward, easily understandable language, with an average FKGL score of 6.44, as shown in Table \ref{5}.

However, the model faces challenges in accurately classifying reviews labeled as “Both”, which contains both FRs as well as NFRs aspects. The F1 score for this category was notably lower, at 0.54. This challenge arises from the inherent complexity of reviews that address both types of requirements simultaneously. The overlap of themes and attributes in such reviews complicates the model’s ability to clearly distinguish between the two. Additionally, reviews in this category tend to exhibit more complex language, as indicated by a higher FKGL score of 7.66. The increased linguistic complexity likely contributes to the model’s reduced classification accuracy in this category.

\begin{table}[htbp]
\centering
\caption{Model performance on different classes}
\centering
\begin{tabular}{|c|c|c|} % 2 columns with vertical lines
\hline
\textbf{Class} & \textbf{F1 Score} & \textbf{FKGL Score} \\ \hline
Functional & 0.91 & 6.55 \\ \hline
Non-Functional & 0.87 & 6.90 \\ \hline
Both & 0.54 & 7.66 \\ \hline
Neither & 0.94 & 6.44 \\ \hline
\end{tabular}
\label{5}
\end{table}

To gain deeper understanding of the model’s performance, we performed a manual analysis of a sample of 100 misclassified reviews. The goal of this analysis was to identify recurring patterns and specific review types where ChatGPT, under the optimized configuration, exhibited misclassification errors. The following sub-sections outline the key issues and case types that contributed to these misclassifications.

\begin{itemize}
    \item \textbf{Negative Sentiment Bias:} One recurring issue observed in the model’s misclassifications is a bias towards negative sentiment. The model tends to prioritize strongly negative reviews, classifying them as indicative of functional issues, regardless of the underlying intent. This bias results in an overrepresentation of functional complaints, while non-functional concerns, often expressed in more moderate language, are underrepresented or overlooked. For example, consider the review: \textit{“The latest update is so frustrating! I can’t believe how bad the app looks now. The new design is horrible, and it’s such an eyesore. I hate using it like this—seriously considering switching to another platform. It’s disappointing.”} While this review primarily critiques the app’s design, a non-functional concern, the emotionally charged language (e.g ‘frustrating’,‘horrible’,‘hate’ and ‘bad’) and intense negative sentiment lead the model to misclassify it as a functional issue.

    \item \textbf{Overlapping Characteristics:} A significant challenge in classifying app review arises when requirements overlap, complicating the model’s ability to accurately identify the intended category. For example, consider the review: \textit{“I used to love this game, but now I’m frustrated. It’s nearly impossible to pass levels without spending money, and the constant pop-ups are annoying. Replaying the same level endlessly is frustrating, and the candies often don’t work properly. It’s disappointing to see a game I once enjoyed become focused on money.”} In this case, the model misclassifies the review as a NFR rather than “Both”. The model emphasizes the mention of “constant popups”, which reflects user experience and usability concerns—categorizing them as non-functional aspects of the game. However, the phrase “the candies often don’t work properly” refers to a functional issue with the game’s mechanics. The model may misclassify such reviews by overemphasizing one aspect (in this case, non-functional) while neglecting the other (functional), resulting in an incomplete classification.

    \item \textbf{Ambiguity in Language:} Ambiguity in language also poses a challenge for accurate classification. Consider the review: \textit{“The app is okay, but the features are confusing. It makes the whole experience frustrating”}. The phrase “features are confusing” can be interpreted in multiple ways: it could refer to a functional issue, such as unclear or poorly designed features, or a non-functional issue, such as poor usability or user interface design. This linguistic ambiguity requires the model to carefully consider contextual cues to accurately distinguish between functional and non-functional concerns. Accurate interpretation of such reviews is crucial for identifying specific areas of improvement, underscoring the importance of context in understanding user feedback.

    \item \textbf{Emotionally Charged Reviews:} Emotionally charged reviews, which emphasize intense emotional reactions or overwhelmingly positive feedback, can often be misclassified during the analysis process. For example, consider the following review, which provides detailed praise for the tracking system but does not highlight any specific functional requirements. Despite the absence of a clear issue, the model erroneously classifies the review as referencing an NFR due to the high level of detail in the description. The review’s emotional tone and focus on user satisfaction may contribute to this misclassification.
    Example: \textit{“I downloaded it about 10 to 15 days ago and I am shocked to see its accuracy... It's just unbelievable. Everyday I travel a huge distance by couple of different vehicles and It not only just tracks the distance but also It tracks by which vehicle and how much time you were on that vehicle too... I am shocked man!!!!”}
\end{itemize}

Table \ref{table6} presents the frequency of misclassifications across the identified error types. The highest misclassification rates were observed in reviews exhibiting overlapping characteristics of both FRs and NFRs, as well as those containing ambiguous content. The “Others” error category included reviews written in informal language, those containing conditional statements, and review featuring complex technical jargon.

\begin{table}[htbp]
\caption{Error categories with their count}
\centering
\begin{tabular}{|c|c|} % 2 columns with vertical lines
\hline
\textbf{Error Category} & \textbf{Frequency} \\ \hline
Negative Sentiment Bias &   19    \\ \hline
Overlapping Characteristics &   33  \\ \hline
Ambiguity in Language &  22 \\ \hline
Emotionally Charged Reviews &  17  \\ \hline
Others &  9 \\ \hline
\end{tabular}
\label{table6}
\end{table}

\section{Threats to validity}

Several potential threats to the validity of our findings must be considered.

The first threat concerns the generalizability of our dataset. While our benchmark includes 1,880 manually annotated reviews, its limited size may not capture the full diversity of language used in app reviews. To address this limitation, we included reviews from ten distinct app domains, ensuring broad linguistic variation and supporting the generalizability of our conclusions.

The second threat concerns annotation bias arising from the subjective nature of classifying app reviews. To mitigate this, five experienced software engineers independently annotated the dataset, with reviews retained based on majority consensus (at least three annotators in agreement), and disagreements resolved through discussion to ensure consistency and accuracy in labeling process.

The third threat concerns prompt engineering, as zero-shot models like ChatGPT are sensitive to prompt structure, with slight wording variations affecting classifications. To minimize this, we tested various prompt techniques from the literature, evaluating both simple and detailed formulations to identify the most effective approach.

% Finally, we acknowledge that misclassification patterns may vary across applications and are not definitive. To minimize this, we used a dataset of reviews from ten different apps across diverse domains, capturing a broad range of user issues and review ambiguities to avoid app-specific biases.
Finally, we acknowledge that misclassification patterns may vary across applications and models, and are not definitive. To minimize app-specific bias, we used reviews from ten diverse apps to capture a broad range of user issues and linguistic ambiguities. However, some patterns may also reflect model-specific biases—such as an overemphasis on negative sentiment—which can differ across LLMs.

\section{Related Work}
App review classification has been extensively studied, identifying categories like bug reports, feature requests, user experiences, advertisements, performance feedback, and user interface issues \cite{rw17} \cite{rw20}. These categories address software issues, capture user sentiments, and offer suggestions. However, a more focused approach is needed to analyze software's functional and quality aspects. Thus, we focus on studies that classify app reviews into FRs and NFRs using ML and NLP techniques.

Hadi et al.\cite{rw20} used ML techniques such as SVM, SGD, and Random Forest with NLP methods like TF-IDF to classify app reviews into FRs and NFRs. Similarly, Yang et al.\cite{rw2} employed TF-IDF and human-assisted keyword selection to achieve stable precision and recall in classifying app reviews into FRs and NFRs. Wang et al.\cite{rw3} proposed AUG-AC, which combined app changelogs and user reviews to improve classification (FR \& NFR) accuracy. Memon et al.\cite{rw4} introduced a framework with feature extraction, sentiment analysis, and topic modeling using traditional ML methods. Deocadez et al.\cite{rw5} used semi-supervised classification (SSC) with self-labeling algorithms to automate classification, achieving 76\% accuracy with a small labeled dataset. However, a limitation of this approach is that if the algorithm misclassifies instances in the unlabeled dataset, errors can accumulate, leading to reduced accuracy when applied to unseen data. For NFR classification, traditional ML techniques like BoW and TF-IDF were employed in \cite{rw6} and \cite{rw7}, while transformer-based models like BERT and RoBERTa were utilized in \cite{rw8} and \cite{rw9}\ to improve accuracy.

While previous studies rely on traditional ML models and domain-specific labeled datasets, our study introduces a zero-shot learning approach using ChatGPT. This method eliminates the need for domain-specific datasets and enables classification based on the model’s general language understanding, offering a more adaptable and efficient solution for evolving app review datasets.

\section{Conclusions and Future work}

This paper investigates the effectiveness of zero-shot learning for automating the classification of app reviews into FRs and NFRs using the GPT-4o mini model. We evaluated various prompting techniques and design strategies on a dataset of 1,880 reviews manually annotated by five software engineering experts. The results were promising, showing the model's potential to classify app reviews without need for large, domain-specific datasets. We also examined how review length and complexity affected performance and compared the GPT-4o mini model’s effectiveness to traditional ML models. To ensure reproducibility, we have made the source code and annotated dataset publicly available at: \url{https://doi.org/10.5281/zenodo.15036830}

Our findings suggest several avenues for future research. First, further exploration of prompt engineering could enhance the model’s performance, particularly by developing prompts that address challenges identified in our manual analysis, such as ambiguity or complexity in language. Second, we observed that the model frequently misclassified reviews containing complex or difficult-to-read language. Incorporating a review simplification step from our prior work \cite{singh-etal-2024-refining} into the processing pipeline could improve classification accuracy by addressing this issue.

%%
%% The next two lines define the bibliography style to be used, and
%% the bibliography file.
\bibliographystyle{ACM-Reference-Format}
\bibliography{sample-sigconf}

%%
%% The next two lines define the bibliography style to be used, and
%% the bibliography file.

%%
%% If your work has an appendix, this is the place to put it.

\end{document}